\documentstyle[12pt]{article}
\def\etal{{\it et.al.}\/}
\def\ie{{\it i.e.}\/}
\def\etc{{\it etc.}\/} 

\def\eg{{\it e.g.}\/}

\def\twi#1{\widetilde{#1}}

\def\nth#1{{1 \over #1}}
\def\hf{\frac{1}{2}}
\def\roughlyup#1{\mathrel{\raise.3ex\hbox{$\sim$\kern-.75em
\lower1ex\hbox{$#1$}}}}
\def\roughlydown#1{\mathrel{\raise.3ex\hbox{$#1$\kern-.75em
\lower1ex\hbox{$\sim$}}}}

\def\simeq{\roughlyup-}

\def\scrs{\scriptscriptstyle}
\def\ap{\alpha'}
\def\IR{\relax{\rm I\kern-.18em R}}
\def\cam{{\cal M}}

\def\cak{{\cal K}}
\def\GN{G_{\scrs N}}

\def\Sc#1{{{\cal #1}}}
\def\ss#1{{{\scriptscriptstyle #1}}}

\def\eqa{\begin{eqnarray}}
\def\eeqa{\end{eqnarray}}
\def\eq{\begin{equation}}
\def\eeq{\end{equation}}
\def\nn{\nonumber}
\def\pref#1{(\ref{#1})}

\begin{document}


\rightline{McGill-95/44}
\rightline{NEIP-95-009}

\vspace{0.5cm}

\begin{center}
{\bf FOUR DIMENSIONAL BLACK HOLES \\ 
AND DUALITY IN SUPERSTRING THEORY}
\end{center}

\vspace{0.3cm}

\begin{center}
C.P. BURGESS\footnote{Invited talk presented to the Sixth
Canadian Conference on General Relativity and Relativistic Astrophysics,
Fredericton, NB, May 1995.}, R.C. MYERS \\ 
\vspace{0.3cm} 
{\em Physics Department, McGill University \\
3600 University St.,  Montr\'eal, Qu\'ebec, Canada, H3A 2T8.} \\
\vspace{0.3cm}
and \\
\vspace{0.3cm}
F. QUEVEDO\footnote{Address after Sept. 1, 1995: Theory Division, CERN,
CH-1211, Gen\`eve 23, Switzerland.} \\ 
\vspace{0.3cm}
{\em  Institut de Physique, Universit\'e de Neuch\^atel \\
1 Rue A.L. Breguet, CH-2000, Neuch\^atel, Switzerland.} \\
\end{center}

\setlength{\baselineskip}{2.6ex}

{\begin{center} ABSTRACT \end{center}
{\small \hspace*{0.3cm} 
Some recent results on the applications of duality (and related)
transformations to general four-dimensional, spherically symmetric,
asymptotically flat and time-independent string configurations are summarized. 
Two classes of results have been obtained. First, these transformations are
used to generate the general such solution to the lowest-order field equations
in the $\ap$ expansion. Second, the action and implications of duality (based
on time-translation) on the general configuration is determined. It is found to
interchange two pairs of the six parameters which label these configurations,
namely: (1) the mass with the dilaton charge, and (2) the axion charge with the
Taub-NUT parameter. For the special case of the Schwarzshild black hole this
implies the relation $M \to - k/\ap M$, where $k$ is a known, positive,
dimensionless number. It is argued that, in some circumstances, dual theories
need not be equivalent in the simplest sense. }}

\section{Introduction}

This article is meant to describe some preliminary applications of duality to
unravelling the implications of string theory for black hole physics in four
dimensions \cite{bhsoln,bhdual}. Before summarizing the points to be outlined,
some motivation is in order as to why duality transformations, and their
applications to black hole physics, are worth thinking about.

Duality transformations \cite[gives an extensive review]{dualityreview} are a
particular type of change of variable which have come to play increasingly
important roles in extracting the physical content of string theory, and of some
ordinary quantum field theories. Field theories which are related by duality
transformations are thought to be completely equivalent, even though they may
appear to be very different. Since it is sometimes true that intractible
strongly-coupled theories turn out to be dual to calculable weakly-coupled ones,
the physical equivalence of the dual theories can be exploited to infer the
behaviour of otherwise unsolvable systems. 

The development of this new tool may open new approaches to studying old
problems of compelling physical interest which have hitherto been too difficult
to crack. In particular, a current hope is that they may be useful for
extracting the predictions of string theory for the dynamics of spacetimes 
having very strong, or even singular, curvatures, such as are believed to arise
in the final state of runaway gravitational collapse, or in the earliest moments
of the universe. The predictions of string theory are of particular interest
for this long-standing puzzle, since string theory is the only known theory
which gives sensible results for the simpler, but related, problem of the
scattering of particles in flat space at energies at and above the Planck-scale.

The good news (so far) is that there are preliminary indications that string
theory has qualitatively new features which may figure importantly in our final
understanding of gravitational collapse. The main new feature to emerge so far
is the realization that string theories appear to be quite forgiving in their
notion of what constitutes a physically unacceptable singularity. What might
be a malignantly singular field configuration from the point of view of ordinary
point-particle field theories, can be completely benign as a background for
string propagation. This understanding has emerged as more solutions have been
constructed \cite[for a recent review]{exactsolns} to the full string equations,
including some singular ones. It is also indicated by the existence of duality
transformations which relate spacetimes with singularities to duals which are
equivalent, and yet are absolutely nonsingular (including even Minkowski space).
Some of the dual spacetimes which have been related in this way even include
black hole spacetimes \cite{twodbhone,twodbhtwo,twodbhthree}, although
admittedly only in two spacetime dimensions.

All of this motivates further study of black-hole-type field configurations
within the context of string theory. This article reports on the results of two
lines of inquiry in this direction. These two lines may be summarized as
follows: 

\begin{enumerate}

\item
The first investigation \cite{bhsoln} uses duality transformations, and some of
their extensions, to generate new classical solutions to the string equations of
motion, starting from simple initial solutions. In particular, they are used to
systematically construct all possible spherically-symmetric, time-independent
and asymptotically-flat solutions in four dimensions. There are two new features
to these results. ($i$) First, because using duality to generate new solutions
is a purely algebraic proceedure, it is much simpler than a direct attempt to
solve the relevant set of nonlinear, coupled, partial-differential equations.
This simplicity has permitted the inclusion of more nontrivial background
fields than previously had been tractable. ($ii$) Second, keeping in mind the
broad-mindedness of string theory towards singularities, {\em all} solutions are
presented, including some that are singular. Singular solutions were often
omitted in previous constructions.  

\item
The second application \cite{bhdual} is to critically analyze the situation
under which dual configurations can be expected to be physically equivalent. In
particular, an important class of duality transformations --- including most of
those applied to black holes --- are argued to {\em not} relate
physically-equivalent spacetimes in the usual sense. The same considerations
also lead to restrictions on the boundary conditions which must be satisfied by
some fields in order for the duality transformations to take their usual form.
\end{enumerate}

These two developments are presented in the remainder of this article in the
following way. \S 2 sets the stage by briefly presenting a reminder
of the connection between string field configurations, and two-dimensional 
conformal field theories. This is followed by the two applications listed
above. First, \S 3 through \S 5 present the use of duality-related
transformations for constructing solutions to the low-energy field equations.
\S 3 summarizes the transformations themselves, while \S 4 gives 
the simplest --- \ie\ dilaton-metric --- solutions. \S 5 then applies the
transformations of \S 3 to the field configurations of \S 4, thereby
generating solutions also incorporating nonzero axion and gauge fields. 

The second line of argument is the topic of \S 6 through \S 8. \S 6 states
the form of the duality transformations in a manner which is sufficiently
general for the desired applications. Their implications for the general
spherically-symmetric, time-independent and asymptotically-flat solutions,
including in particular the black hole solutions, are the topic of \S 7. The
result, as applied to black hole configurations, makes it difficult to see how
the dual solutions can be physically equivalent. \S 8 is dedicated to the
resolution of this puzzle. 

The conclusions which follow from these sections are finally summarized in \S 9. 

\section{Strings and 2D Field Theories} 

This section is meant to review the connection between string field
configurations, and two-dimensional conformal field theories, since this
underlies all of what follows. This connection allows the application of 
results for two-dimensional field theories to draw conclusions about 
solutions to the string field equations.

\subsection{String Field Configurations}

A classical solution to the string theory equations of motion is equivalent to
a conformally-invariant, two-dimensional field theory. The connection is
simplest to see in the case that the two-dimensional field theory is a sigma
model, whose two-dimensional scalar fields, $x^\mu(\sigma)$, can be interpreted
as describing the worldsheet of a string propagating within a `target'
spacetime. For instance, suppose the sigma model action has the form
\eqa
\label{sigmamodelaction}
S[x^\mu] &=& - \, {1 \over 2 \pi \alpha'} \int d^2\sigma \; \sqrt{\gamma} \,
\Bigl[ G_{\mu\nu}(x) \, \gamma^{\alpha\beta}  + B_{\mu\nu}(x) \,
\epsilon^{\alpha\beta} \Bigr] \; \partial_\alpha x^\mu \partial_\beta x^\nu
\nn\\
&& \qquad \qquad \qquad - \, {1 \over 8 \pi} \int d^2\sigma \; \sqrt{\gamma} 
\; \phi(x) \Sc{R} , 
\eeqa
where $\gamma_{\alpha\beta}$ and $\epsilon^{\alpha\beta}$ are the
two-dimensional metric and antisymmetric Levi-Civita symbol and $\Sc{R}$ is the
curvature scalar for the metric $\gamma_{\alpha\beta}$. 

The coupling functions, $G_{\mu\nu}(x)$, $B_{\mu\nu}(x)$ and $\phi(x)$, are
interpreted in string theory as being background values for three fields ---
the metric, the `axion', and the `dilaton' --- which represent three of the
modes of the string. These particular modes would have been massless if the
string were propagating through Minkowski space. There are typically other such
nearly `massless' modes as well in a string theory, and we shall consider later a
spin-one gauge fields, $A_\mu(x)$, in addition to the above three. 

Now comes the main point. There are two ways in which one might imagine
defining the equations of motion for these fields, given our presently limited
understanding of string theory itself. The direct way is to infer the
interactions of the various string modes by computing their tree-level
scattering on simple spacetimes, such as for Minkowski space. One finds an
action which reproduces these scattering amplitudes, and then computes the
equation of motion using this action.

The alternative way to determine the equations of motion for $G_{\mu\nu}$,
$B_{\mu\nu}$, $\phi$ and $A_\mu$ is to compute the conditions which these
quantities must satisfy in order for the corresponding two-dimensional theory to
be conformally invariant at the quantum level. It is an amazing fact that these
two methods produce results which agree with one another --- up to the
ubiquitous freedom to perform field redefinitions --- in all cases for which
they have been compared. 

The resulting field equations can be explicitly written as an expansion in
powers of derivatives of the background fields times the dimensionful constant
$\alpha'$. In four dimensions the action which reproduces the leading terms in
these equations for the fields of interest is \cite{GSW}:  
\eq
\label{leaction}
\Sc{L} = {e^\phi  \over 8 \pi \ap} \; \sqrt{- G} \, \left[ R(G) + (\nabla
\phi)^2 - {1 \over 12} \, H^{\mu\nu\lambda} H_{\mu\nu\lambda} - {1\over 8} \,
F^{\mu\nu}F_{\mu\nu} \right] + \cdots \, , 
\eeq
where $H_{\mu\nu\lambda} = \partial_\mu B_{\nu\lambda} - \nth{4} \, A_\mu
F_{\nu\lambda} + \hbox{(cyclic permutations)}$\footnote{By including the
gauge-field-dependent Chern-Simons term, $\nth{4} \, AF$, in the definition of
$H$ we restrict our discussion of gauge fields to the heterotic string. By
contrast, the discussion for $A_\mu = 0$ applies equally well to bosonic
and superstrings.} and $F_{\mu\nu} = \partial_\mu A_\nu - \partial_\nu A_\mu$
are, respectively, the field strengths for the axion and electromagnetic fields.
$R(G)$ is the Ricci scalar for the metric, $G_{\mu\nu}$, and the square root
involves the quantity: $G = \det G_{\mu\nu}$. This metric is often called the
`sigma-model' metric to distinguish it from the one for which a field
redefinition has been performed to put the Einstein term into standard form. The
ellipses in eq.~\pref{leaction} represent terms which involve other massless
fields and others involving more derivatives that arise at higher orders in
$\alpha'$. 

Some of the higher-order corrections to these equations in the derivative
expansion have also been computed. For later purposes we quote the quantity
which is responsible for the higher-order corrections to configurations
involving only the metric and the dilaton, which turns out to be \cite{GSW}: 
\eq
\label{corrtoleaction}
\delta \Sc{L} = {\lambda e^\phi \over 2} \; \sqrt{-G} \; 
R_{\mu\nu\rho\sigma}R^{\mu\nu\rho\sigma} \ .
\eeq
Here $\lambda$ is $\hf$ for bosonic strings, $\nth{4}$ for heterotic strings,
and $0$ for supersymmetric strings. 

This establishes a direct connection between background string configurations
and two-dimensional conformal field theories. The spectrum of string states
corresponds to a class of operators in these field theories which represent the
conformal group in a particular way. The scattering of these string states can
be computed by evaluating appropriate correlation functions involving these
operators in the conformal field theory. 

Given this connection it is immediate how to interpret the equivalence between
any two conformal field theories in terms of string states. If two conformal
theories can be shown to have precisely the same content, then they must
describe identical string scattering about identical background string
configurations. Duality transformations are interesting because they imply
precisely this kind of equivalence between two-dimensional field theories. 

\section{Classical Transformations}

There is a broad class of transformations which are guaranteed to map conformal
field theories into other conformal field theories, even though the theories
which are related in this way need not be physically equivalent. That is to
say, this broader class of transformations are guaranteed to take classical
string vacua into other classical vacua, but the full Hilbert space of string
modes constructed about these vacua can be different. Although these
transformations cannot therefore be considered to be {\it bona-fide} string
symmetries, they are nevertheless very useful for generating new classical
solutions from known ones. 

There are two transformations of this type which are used in what follows. We
outline both of these in the following two subsections.

\subsection{$SL(2,\IR)$ Transformations}

The first class of such transformations is a group of $SL(2,\IR)$
transformations \cite{STW,MDR,Sen} which includes the classical string
$S$-duality transformations \cite{Wit,BFQ,LPS} of the low-energy effective
theory. To formulate these transformations it is useful to use the scalar
variable, $a(x)$, which is dual to the three-index axion field,
$H_{\mu\nu\lambda}$: 
\eq \label{axion}
H_{\mu\nu\rho} = -e^{-2\phi}\epsilon_{\mu\nu\rho\kappa}\nabla^\kappa a .
\eeq
Here all indices are raised and lowered with the Einstein metric,
$g_{\mu\nu}$, which (in four spacetime dimensions) is related to the sigma-model
metric, $G_{\mu\nu}$ by: $g_{\mu\nu} = e^\phi \; G_{\mu\nu}$, and the
Levi-Civita tensor is also constructed using $g_{\mu\nu}$. Many properties of
the theory take a simple form if the field, $a$, is combined with the dilaton,
$\phi$, into the complex combination, $S = a + i\, e^{\phi}$. In terms of this
variable the $SL(2,\IR)$ transformation becomes
\eqa
\label{sltrt}
S &\longrightarrow& { {\rm a}\, S + {\rm b} \over {\rm c} \,
S +{\rm d}} \nn\\
(F_+)_{\mu\nu} &\longrightarrow& \left( {\rm c}\, S +{\rm d} \right)\,
\left( F_+ \right)_{\mu\nu} \\
(F_-)_{\mu\nu} &\longrightarrow& \left( {\rm c}\,  S^* +{\rm d} \right)\,
\left( F_- \right)_{\mu\nu} . \nn
\eeqa
where $S^*$ is the complex conjugate of $S$, and $(F_\pm)_{\mu\nu} \equiv
F_{\mu\nu} \pm {i\over2} \; \epsilon_{\mu\nu\rho\kappa} \, F^{\rho\kappa}$.
Once again it is the Einstein metric which is involved in these definitions. The
quantities a, b, c and d are real numbers which must satisfy ${\rm ad} - {\rm
bc} = 1$. If $S$, $g_{\mu\nu}$, and $F_{\mu\nu}$ all satisfy the string
equations of motion, then so must the transformed variables as defined by
eq.~\pref{sltrt}.

\subsection{The $O(1,1)$ Transformations}

There is a second transformation which can be used to generate new classical
string solutions from old ones \cite{CFG,MV,Sen92,GMV,HS}. In the simplest case
of a field configuration that is independent of a single coordinate, $s$, there
is an $O(1,1)$ group of such transformations. The action of these
transformations is most easily written when the background fields are written as
the following $9 \times 9$ matrix  
\eq
\label{Mmatrix}
{\cal M} =
\pmatrix{\cak_-^TG^{-1} \cak_- & \cak_-^TG^{-1}\cak_+ & -\cak_-^TG^{-1}A \cr
       \cak_+^TG^{-1}\cak_- & \cak_+^TG^{-1}\cak_+ & -\cak_+^TG^{-1}A \cr
       -A^TG^{-1}\cak_- & -A^TG^{-1}\cak_+ & A^TG^{-1}A \cr}
\eeq
where
\eq
\label{Kmatrix}
(\cak_\pm)_{\mu\nu}=-B_{\mu\nu}-\Sc{G}_{\mu\nu} \pm \eta_{\mu\nu} ,
\eeq
and $\eta_{\mu\nu}$ is the flat Minkowski metric in four dimensions. In these
expressions the quantity $\Sc{G}_{\mu\nu}$ is defined by: $\Sc{G}_{\mu\nu} =
G_{\mu\nu} + \nth{4} \, A_\mu A_\nu$. For a detailed statement of the
conventions used, see ref. \cite{bhsoln}.

With these variables the $O(1,1)$ transformations can be expressed in matrix
form, $\cam\to \Omega \cam\Omega^T$, where the transformation matrix is
given by 
\eq
\label{Omatrix}
\Omega=\pmatrix{I_7&0&0\cr0&x&
\sqrt{x^2-1}\cr0&\sqrt{x^2-1}&x} .
\eeq
Here $I_7$ is the $7\times7$ unit matrix, and $x$ is a parameter satisfying
$x^2 \geq 1$. The action of the $O(1,1)$ transformations on the dilaton is
given by 
\eq
\label{dilatonp}
e^{\phi} \to \left({{\rm det}G\over{\rm
det}G^\prime}\right)^{1\over2} e^\phi\ \ ,
\eeq
where $G'_{\mu\nu}$ denotes the transformed sigma-model metric. 

\section{Dilaton-Metric Configurations}

The goal now is to use these transformations to generate the most general
spherically-symmetric, asymptotically-flat and static solutions to the string
equations. This will be done by applying the transformations to a particularly
simple class of solutions. The first step --- identifying this initial simple
class of solutions --- is the topic of the present section.

\subsection{The Lowest-Order Solution}

The starting point is the most general four-dimensional, asymptotically flat,
static and spherically-symmetric configuration which solves the string
equations to leading order in $\ap$, and which involves only the metric and
dilaton fields.  We therefore directly solve the equations: 
\eqa
\label{eqofmotion}
R_{\mu\nu}(g) &=& \hf \nabla_\mu \phi \nabla_\nu \phi \nn\\
\nabla^2 \phi &=& 0 ,
\eeqa
using the {\em ansatz}
\eqa
\label{sssconfig}
g_{\mu\nu} dx^\mu dx^\nu &=& - f(r) dt^2 + g(r) dr^2 + h^2(r) (
d\theta^2 + \sin^2 \theta \, d\varphi^2 ) , \nn\\
\phi &=& \phi(r),  \qquad 
\eeqa

The result is a family of field configurations whose study actually predates
the earliest advent of string theory itself \cite{Buchdahl,JRW}:
\eqa
\label{answer}
f = {1 \over g} &=& \Lambda^\delta(r) \nn\\
h^2 &=& r^2 \Lambda^{1-\delta}(r) \\
e^\phi &=& e^{\phi_0} \; \Lambda^\gamma(r) , \nn
\eeqa
where $\Lambda(r)$ is a shorthand for the function $1 - (\ell / r)$. $\ell$ is
the only dimensionful parameter in the solutions, and so it simply sets their
overall scale. We assume both $\ell$ and $r$ to be positive in what follows.
The two dimensionless quantities, $\delta$ and $\gamma$, are the arbitrary
parameters which label the solutions, subject to the condition $\delta^2
+ \gamma^2 = 1$. 

$\phi_0$ is the asymptotic value which is obtained by the dilaton field as $r
\to \infty$. It is convenient in what follows to shift $\phi$ by a constant to
ensure that it tends to zero for large $r$, and so to completely remove $\phi_0$
from the solutions. This is always possible in string theory, with the general
result --- for classical string solutions --- that $\phi_0$ only appears as an
overall factor in the low-energy lagrangian of eqs.~\pref{leaction} and
\pref{corrtoleaction}. It can therefore be absorbed into the definition of
Newton's constant, which we take (in four dimensions) to be $\GN = \hf \,
e^{-\phi_0} \, \ap$. 

The choice $(\delta,\gamma)=(1,0)$ yields the Schwarzschild metric with constant
dilaton field, and $\ell$ is in this case related to Newton's constant, $\GN$,
and the black hole mass, $M$, by $\ell = 2 \GN M$. So long as $\gamma \ell \neq
0$, however, the metric of eq.~\pref{answer} has a real curvature singularity at
$r=\ell$.

\subsection{The Conserved Charges}

The two independent parameters which label this family of solutions have a
useful interpretation in terms of the asymptotic behaviour, as $r \to \infty$,
of the fields involved. In general, for large $r$, the solutions take the form:
\eqa
\label{gasymform}
f(r) &=& 1 - {\Sc{A} \over r} + \cdots;  \nn\\
g(r) &=& 1 + {\Sc{B} \over r} + \cdots ; \\
h(r) &=& r^2 \left[ 1 + O\!\left( {1 \over r} \right) \right] \, , \nn\\
e^{\phi(r)} &=& 1 - {Q_{\ss{D}} \over r} + \cdots \, , \nn
\eeqa
with constants, $\Sc{A} = \Sc{B} = \delta \ell$ and $Q_{\ss{D}} = \gamma
\ell$. 

These constants have a physical interpretation as specifying the corresponding
conserved charges which are carried by the solutions. That is, the conserved
(ADM \cite{ADM}) mass, $M$, of the solution is related to the constant $\Sc{A}$
by $2 \GN \, M = \Sc{A}$. The constant, $Q_{\ss{D}}$, similarly defines a {\em
dilaton charge} for the solution. 

The utility of labelling the solutions by their values for $M$ and $Q_{\ss{D}}$
is that these quantities are equally well defined for the complete solution to
the string equations. Given these asymptotic expressions, the complete equations
may be solved order by order in $\ap$, giving a unique solution for {\em
any} choice of $M$ and $Q_{\ss{D}}$. The same is not true for the parameters
$\delta$ and $\gamma$, which are defined in terms of the specific form of the
solution to these equations only at lowest order in $\ap$.  For instance, if we
focus on the black hole solution --- which we take to be the solution for which
the potential singularity at $r = \ell$ is only a coordinate artifact --- and
work to next order in $\ap$ by including eq.~\pref{corrtoleaction} into the
low-energy string action, then the result is still characterized by the
quantities $M$ and $Q_{\ss{D}}$, but with \cite{CMP}:   
\eqa
\label{bhasymform}
2 \GN M = \Sc{A} = \Sc{B} &=& \ell + {11 \lambda \ap  \over 6 \ell} 
+ O(\ap^2) ,\nn\\
Q_{\ss{D}} &=& - \; {2 \lambda \ap  \over \ell} + O(\ap^2) .
\eeqa
Recall that the constant $\lambda$ appearing here is $\hf$ in the bosonic
string, $\nth{4}$ in the heterotic string, and 0 in the superstring.

\section{More General Classical Solutions}

It is now possible to compute more general classical solutions, simply by
repeatedly applying the $SL(2,\IR)$ and $O(1,1)$ transformations of the
previous sections to the above solutions. Before doing so explicitly, it is
first worth characterizing the possible parameters which can describe these
solutions by first analyzing the asymptotic behaviour which is permitted for
the fields at large $r$. 

\subsection{The Conserved Charges}

Once the metric and dilaton fields are supplemented by nonzero axion and gauge
fields, more complicated solutions become possible. We take the following
spherically-symmetric and time-independent {\em ansatz} for the axion
and gauge fields: 
\eq
\label{sssdilem}
a = a(r), \qquad F_{tr} = E(r), \qquad F_{\theta \varphi} = B(r) \sin \theta.
\eeq
and we generalize, for later convenience, our metric {\em ansatz} to include
stationary, but not static, metrics:
\eq
\label{taubnutansatz}
g_{\mu\nu} dx^\mu dx^\nu = - f(r) (dt + 2 N \cos\theta \; d\phi)^2 + g(r) dr^2 +
h^2(r) ( d\theta^2 + \sin^2 \theta \, d\varphi^2 ).
\eeq
The parameter $N$ here is called the {\em NUT} parameter, due to the
similarity of eq.~\pref{taubnutansatz} with the Taub-NUT metric
\cite{Taub,NUT,CEtoE}. 

The corresponding new conserved charges can be inferred by examining the 
large-$r$ behaviour that is implied by the field equations for these fields.
This is given, for the dilaton and metric, by eqs.~\pref{gasymform}, and for the
axion and gauge fields by:
\eqa
\label{emaform}
a(r) &=& - \; {Q_{\ss{A}} \over r} + \cdots; \nn\\
A_t &=& {Q_{\ss{E}} \over r} + \cdots, \\
A_{\varphi} &=& -\; Q_{\ss{M}}\, \cos\theta + \cdots .\nn
\eeqa
The constants $Q_{\ss{A}}$, $Q_{\ss{E}}$ and $Q_{\ss{M}}$ are the solution's
axion, electric and magnetic charges, respectively. 

\subsection{Dilaton-Axion-Metric Solutions}

The simplest generalization is the inclusion of a nonzero axion field in
addition to the original dilaton and metric configurations. These may be
generated by applying the $SL(2,\IR)$ transformations, eq.~\pref{sltrt}, to
the solution of eq.~\pref{answer}, being careful to preserve the boundary
conditions for $\phi$ and to ensure that $a \to 0$ at large $r$. (There is no
loss of generality in choosing this boundary condition for the axion field,
since the definition, eq.~\pref{axion}, only defines $a(r)$ up to an additive
constant.)

Performing the $SL(2,\IR)$ transformation, we obtain the same metric
configuration as before, but the new dilaton and axion fields, $\hat\phi$ and
$\hat a$:  
\eq
\label{firstt}
e^{\hat\phi} = \left(1+\omega^2\right){\Lambda^\gamma(r)
\over \omega^2 \Lambda^{2\gamma}(r) +1} , \qquad
\hat a ={\omega \left[ \Lambda^{2\gamma}(r) -1 \right] \over
\omega^2 \Lambda^{2\gamma}(r) +1} ,
\eeq

$\omega$ is the new real parameter of the solution, while $\delta$ and $\gamma$
are the labels of the original dilaton-metric configuration. The starting
dilaton-metric solution is re-obtained as the special case $\omega = 0$. These
parameters are related to the three conserved charges of the
dilaton-axion-metric system by: 
\eq
\label{charges}
2 \GN \, M = \delta \ell , \qquad
Q_{\ss{D}}= {1-\omega^2 \over 1+\omega^2}\; \gamma \, \ell \qquad
\hbox{ and} \qquad Q_{\ss{A}} ={2\omega \, \gamma \, \ell\over 1+\omega^2} .
\eeq

All of these solutions have real singularities at $r=\ell$, except for the
special case $\gamma \ell =0$. An interesting special limit of these solutions
is  the case $\omega=\pm1$, for which the dilaton charge, $Q_{\ss{D}}$, vanishes.
Notice that even for this choice, however, the dilaton field is nonvanishing
due to the nontrivial axion configuration. 

\subsection{Including a Gauge Potential}

It is conceptually no more difficult, although algebraically more tedious, to
generate the general dilaton--axion--metric--gauge-potential configuration.
This is obtained by repeatedly performing $SL(2,\IR)$ and $O(1,1)$
transformations to the previous solutions. Three new parameters, $x$, $\epsilon$
and $\rho$, enter the solution in this way before these transformations start
just regenerating previously-obtained configurations. These three parameters
are related to the electric and magnetic charges, $Q_{\ss{E}}$ and $Q_{\ss{M}}$,
of the background electromagnetic fields, and to the NUT parameter, $N$,
of the metric (which is defined by eq.~\pref{taubnutansatz}). 

The result of this process is a fairly complicated field configuration, whose
explicit form is not particularly illuminating and so is not repeated here.
Detailed expressions are given in \cite{bhsoln}. Instead, we turn to the action
of duality on these solutions, and on their extensions to higher order in the
$\ap$ expansion.

\section{Duality Transformations}

The remainder of this article is devoted to exploring the implications of
duality transformations for the black-hole, and singular, solutions just
constructed. Some of the results we obtain also apply to the exact solutions
to the string equations. 

\subsection{The Transformation Rules}

The first step is to define what is meant by duality transformations.
A reasonably general algorithm has emerged with which it is always possible to
systematically generate dual field theories from a given one
\cite{algorithm,AAL,AABL}, generalizing an earlier construction which had
been developed for the earliest known string duality \cite{Buscher}. The
algorithm applies to any field theory which admits a continuous symmetry. 

It is simple to state the result for the case of the sigma model,
eq.~\pref{sigmamodelaction}, when the symmetry corresponds to the independence
of one of the coordinate directions, of the fields $G_{\mu\nu}$, $B_{\mu\nu}$,
$\phi$ and $A_\mu$. Denoting this direction by $s$, then the sigma model which
is dual to the original one involves the fields $\twi{G}_{\mu\nu}$,
$\twi{B}_{\mu\nu}$, $\tilde\phi$ and $\twi{A}_\mu$, where
\cite{Buscher,GRV,SW,AAB}:   
\begin{eqnarray}
\label{dualitytransformation}
\twi{\Sc{G}}_{ss} &=& 1/\Sc{G}_{ss},\qquad
        \twi{\Sc{G}}_{s\mu}=-B_{s\mu}/\Sc{G}_{ss},\nn\\
 \twi{\Sc{G}}_{\mu\nu} &=& \Sc{G}_{\mu\nu} - \left[ 
{\Sc{G}_{s\mu}\Sc{G}_{s\nu} + B_{s\mu} B_{s\nu}\over \Sc{G}_{ss}} \right] \\
\twi{B}_{s\mu} &=& -{\Sc{G}_{s\mu}}/{\Sc{G}_{ss}},\qquad
       \twi{B}_{\mu\nu}=B_{\mu\nu}+{\Sc{G}_{s\mu} B_{s\nu}
         -\Sc{G}_{s\nu}B_{s\mu}\over \Sc{G}_{ss}}, \nn \\
\twi{A}_s &=& - \; {A_s \over \Sc{G}_{ss}} \nn\\
\twi{A}_\mu &=& A_\mu -A_s \; {\Sc{G}_{s\mu} -B_{s\mu} \over \Sc{G}_{ss}} \nn\\ 
e^{\tilde\phi} &=& e^\phi\left({\det G\over \det\twi G}\right)^{1/2} .\nn
\end{eqnarray}
As before, the quantity $\Sc{G}_{\mu\nu}$ is defined by:
$\Sc{G}_{\mu\nu} = G_{\mu\nu} + \nth{4} \, A_\mu A_\nu$, and the index $\mu$
runs over all values except for $\mu = s$. (Ref. \cite{AABL} presents the
duality transformations in a more manifestly covariant way.) These
transformations in general can acquire higher-derivative corrections in powers
of $\ap$ as well. 

\subsection{An Example: The Torus}

Perhaps the simplest example brings out many of the main features of these
duality transformations.\footnote{See ref.~\cite{dualityreview} for references
to the extensive literature on toroidal duality.} The simplest example consists
of a toroidal spacetime, for which the metric is flat and all other background
fields are taken to be trivial. If we base the duality on the symmetry of
translations along one of the compact coordinate directions of the torus ---
call it $s$, say --- then the general expression of
eq.~\pref{dualitytransformation} simply reduces to the replacement:  
\eq
\label{toruscase}
G_{ss} \to {1 \over G_{ss}}, \qquad \phi_0 \to \phi_0 + \log \Bigl| G_{ss}
\Bigr|. 
\eeq
Here both $G_{ss}$ and $\phi_0$ are constants. The dual configuration is once
more a torus, but eq.~\pref{toruscase} states that if the circumference in the
$s$ direction is initially $R$, then for the dual theory it becomes $\twi{R}
\propto \ap^2/R$. 

The beauty of the toroidal example is that for this background the complete
spectrum of string fluctuations is known, and the correspondence of states in
the dual theories can be explicitly followed. States turn out to be labelled
by two integers, $m$ and $n$, in addition to other quantum numbers, where $m$
labels the quantized momentum in the $s$ direction and $n$ gives the number of
times which the string winds around this direction. It turns out that duality
acts to interchange $m$ and $n$, while leaving all of the other quantum numbers
fixed. The important role which is played by the winding modes, for which $n
\neq 0$, emphasizes the intrinsically `stringy' nature of the duality symmetry.

When the direction associated with the symmetry coordinate is compact, such as
for the translations of a torus just considered, then the sigma models having
the dual field configurations can be shown to be completely equivalent, and so
describe {\em exactly} the same string physics \cite{algorithm}.
Eqs.~\pref{dualitytransformation} can therefore be considered as full quantum
symmetries of string theory. 

The relation of the dual theories is less clear when the relevant symmetry
coordinate is not compact. We argue in what follows that the equivalence need
{\em not} be true in the noncompact case. 

\section{Applications of Duality}

We now explore the implications of duality for general spherically-symmetric,
time-independent and asymptotically-flat field configurations.

\subsection{The Dilaton-Metric Configuration}

To get an idea for what is implied by a duality transformation, we first apply
one to the dilaton-metric configuration considered in \S 4.1, above. We
base the duality transformation on the symmetry of time translation of the
original solution. This solution, given by eq.~\pref{answer}, is characterized
by the two constants, $\delta$ and $\gamma$, together with the dimensional
quantity $\ell$. Applying the duality transformation,
eq.~\pref{dualitytransformation}, to this configuration leads to another
solution of the form as eq.~\pref{answer}, but with the constants $\delta$ and
$\gamma$ interchanged:  
\eq
\label{dualityondilmet}
\delta \longleftrightarrow \gamma. 
\eeq

Given the relation between these parameters and the conserved charges we see
that eq.~\pref{dualityondilmet} implies:
\eq
\label{dualityondilmetchg}
2 \GN \, M \longleftrightarrow Q_{\ss{D}}. 
\eeq
As will now be discussed, this last way of writing the duality transformation
law applies equally well to the solutions at higher-orders in $\ap$. 

\subsection{Application to General Configurations}

Since part of the promise of studying duality transformations lies in their
potential for leading to exact information concerning the theory involved, it
is of enormous interest to understand how duality acts on the exact string
solutions, rather than simply having its action on the approximate solutions
to low orders in the derivative expansion. The obvious difficulty lies in
determining this action when explicit expressions for the solutions, and the
transformation laws, are themselves not known beyond the leading order in
$\ap$.  

It is nonetheless possible to draw general conclusions for the case of
time-independent, spherically-symmetric, asymptotically flat field
configurations, since these are in principle completely characterized by their
values for the conserved charges $M$, $N$, $Q_{\ss{D}}$, $Q_{\ss{A}}$,
$Q_{\ss{E}}$, and $Q_{\ss{M}}$. Furthermore, these charges are themselves
completely determined by the behaviour of the fields in the asymptotic,
large-$r$, regime for which all fields are very slowly-varying. As a result,
it is possible to determine the action of duality on the conserved charges of a
solution using just the leading expressions in the derivative expansion. 

By applying the transformation law, eq.~\pref{dualitytransformation}, to the
asymptotic forms for the background fields, eqs.~\pref{gasymform} and
\pref{emaform}, the action of duality on the space of four-dimensional,
asymptotically-flat, time-independent and spherically-symmetric string solutions 
may be determined in general. We find that the solution which is labelled by
the charges $M$, $N$, $Q_{\ss{D}}$, $Q_{\ss{A}}$, $Q_{\ss{E}}$, and $Q_{\ss{M}}$
becomes mapped to the solution whose charges are $\twi M$, $\twi N$,
$\twi Q_{\ss{D}}$, $\twi Q_{\ss{A}}$, $\twi Q_{\ss{E}}$, and $\twi Q_{\ss{M}}$,
where:
\eq
\label{dualsum}
2 \GN \, \twi{M} = Q_{\ss{D}}, \qquad \twi{Q}_{\ss{D}} = 2 \GN \, M, \qquad
\twi{Q}_{\ss{A}} = 2 N, \qquad 2 \twi{N} = Q_{\ss{A}},
\eeq
and 
\eq
\label{dualnochange}
\twi{Q}_{\ss{E}} = Q_{\ss{E}}, \qquad \twi{Q}_{\ss{M}} = Q_{\ss{M}}.
\eeq
That is, duality simply interchanges the mass with the dilaton charge, as well
as interchanging the axion charge with the NUT parameter, while leaving
the electric and magnetic charges untouched.

\subsection{The Black Hole Special Case}

This transformation law has some odd consequences, as may be seen by focussing
on the black-hole solutions. At lowest order in $\ap$ these are characterized by 
the parameters $\delta = 1$ and $\gamma = 0$, and so their image under duality
must have $\tilde\delta = 0$ and $\tilde\gamma = 1$. If we use the expressions
for $M$ and $Q_{\ss{D}}$, as given to $O(\ap)$ in eqs.~\pref{bhasymform}, then
we find for the dual configuration:
\eqa
\label{smdasymform}
2 \GN \twi M &=& - \; {2 \lambda \ap  \over \ell} + O(\ap^2) , \nn\\
\twi Q_{\ss{D}} &=& \ell + {11 \lambda \ap  \over 6 \ell} + O(\ap^2) . 
\eeqa

This transformation rule contains the seeds of a puzzle. Notice, in this regard,
the mapping from positive to negative mass: 
\eq
\label{mdual}
\twi{M} = - \, {k \over \ap M} + O\!\left({1\over\ap^2M^3}\right),
\eeq
where $k = \lambda \ap^2 / 2 \GN^2 = 2\lambda e^{2\phi_0}$ is a dimensionless
constant. (In the special case of the superstring, for which $\lambda = 0$, it
is necessary to work to still higher order in $\ap$ to infer the sign of
$\twi{M}$, again giving the result \cite{bhdual} that $M$ and $\twi{M}$ have
opposite signs.) The puzzle is to understand how the physical equivalence of
dual solutions can be reconciled with a change of sign of the solution's mass.
The next section is devoted to the resolution of this puzzle.  

\section{Equivalence Under Duality}

Intuitively, it seems absurd that a background field configuration having
negative mass could be physically equivalent to one for which the mass is
positive. The goal of this section is to pin down this intuition, in order to
better understand the circumstances under which dual spacetimes can be 
equivalent. 

A more precise reason for doubting the equivalence of spacetimes having opposite
masses would be the expectation that such spacetimes should gravitationally
scatter incident test particles differently. After all, based on experience
with the Newtonian force law for universal gravitation, a positive mass should
exert an attractive gravitational force on a distant, slowly-moving particle,
while the force due to a negative mass should be repulsive in this limit. One
might worry, however, whether the presence of other massless background fields
--- such as the dilaton or axion \etc\ --- may invalidate any intuition which
is based on the Newtonian limit of pure gravity. 

\subsection{Test-Particle Scattering}

It is straightforward to test these ideas, by computing the scattering of test
particles which remain at large $r$, and so which experience only slowly varying
fields. This calculation may be performed explicitly in string theory by
choosing massless string states as test particles. The result is particularly
simple in the limit for which the wavelength of the incident state is both
longer than the Planck length, and shorter than the typical distances over which
the background fields vary. This is the regime of geometric optics, for which
simple arguments show that massless string states simply follow the null
geodesics of the metric, regardless of the presence of axion and dilaton fields
\cite{bhdual}. 

For example, in this limit the scattering angle, $\Delta \varphi$, of a photon
which remains at large $r$ throughout its interaction with the background
field is given (for vanishing NUT parameter) by \cite{bhdual}
\eq
\label{simpang}
\Delta \varphi \simeq {4 \GN \, M \over r_0} ,
\eeq
where $r_0$ is the smallest radial coordinate that is attained by the photon.
As advertised, this result depends only on the mass $M$ of the source, and not
on its other charges. The result for the dual spacetime is given by the same
expression, with $M \to \twi{M}$.\footnote{This assumes only that the dual
NUT parameter also vanishes, which implies the vanishing of the axion
charge, $Q_{\ss{A}}$, of the original solution. This class of configurations
is amply big for the present purposes.} 

\subsection{A Better Argument}

Convincing as the previous calculation may seem, it leaves room for doubt
concerning the inequivalence of the dual field configurations we have
considered. This is because it leaves open a loophole, whose existence becomes
clear once the well-understood example of toroidal duality is reconsidered. As
was stated in \S 6.2, duality acts to interchange the labels $m$ and $n$
which respectively characterize the momentum and winding number along the
symmetry direction. It is the scattering of `winding modes' ($n \neq 0$) in the
dual theory which is equal to the scattering of `momentum modes' ($m \neq 0$)
in the original theory. The problem with the scattering calculation of \S 8.1
arises because it is not clear whether the photon in the original theory is
mapped to the same photon in the dual theory. If not there is no reason for a
result like eq.~\pref{simpang} to be the same for both the original spacetime
and its dual.

In fact, since the symmetry on which duality is based for the configurations
considered here is time translation, it is not completely clear precisely how
the winding and momentum modes should be defined. As a result, an alternative
argument in favour of the inequivalence of the dual solutions is now presented
which does not rely on being able to trace the detailed action of duality on
the various string states. This is done by identifying a one-parameter family
of equivalent background field configurations, and showing that these dualize
to a one-parameter family of inequivalent configurations. We are led to the
conclusion that eqs.~\pref{dualitytransformation} need not be a string
symmetry, at least when the coordinate `$s$' labels a noncompact direction.

The key point on which this argument relies is the observation that the
transformations of eqs.~\pref{dualitytransformation} depend in detail upon the
boundary condition which is satisfied at large $r$ by the gauge potential in the
time direction, $A_t$. In particular, the duality transformations obtained in
\S 7.2 for the conserved charges of the spherically-symmetric solutions are
only correct if it is assumed that $A_t$ falls off like $1 / r$. This is
required if the quantity $\Sc{G}_{tt} = G_{tt} + \nth{4} \, A_t A_t$ is to have
the same asymptotic form as does $G_{tt}$ --- a result which is required in
deriving eqs.~\pref{dualsum} and \pref{dualnochange}. 

Suppose, then, we instead assume the following asymptotic behaviour for $A_t$:
\eq
\label{newgauge}
A_t=2v+{Q_{\ss{E}} \over r}+\cdots,
\eeq
where $v$ is a constant. This form does not affect the asymptotic behavior of
the electric and magnetic fields, and so does not change the values of the
electric and magnetic charges. In order to not change the asymptotic form for
the field strength, $H=dB-{1\over4}A\,dA$, it is also necessary to alter the
asymptotic behavior of the Kalb-Ramond field, $B$, which we assume to have done
in what follows.

When performing the duality transformation, eq.~\pref{dualitytransformation},
using the asymptotic form of eq.~\pref{newgauge} it is necessary to rescale the
time coordinate to preserve the limit $G_{tt} \to -1$ as $r \to \infty$, and to
shift the dilaton by a constant in order to recover its previous limiting value
at infinity. With these choices the action of duality on the asymptotic charges
is no longer given by eqs.~\pref{dualsum} and \pref{dualnochange}, but rather by
\cite{bhdual}: 
\eqa
\label{longeq}
2{\twi{\GN}} \twi{M} &=& {1\over1-v^2} \left[ Q_{\ss{D}} -2v^2 \GN M -v
Q_{\ss{E}} \right] \nn\\
\twi{Q}_{\ss{D}} &=& {1\over1-v^2} \left[ 2\GN M -v^2 Q_{\ss{D}} +v
Q_{\ss{E}} \right] \nn\\
2\twi{N} &=& {1\over1-v^2} \left[ Q_{\ss{A}} - 2v^2 N \right] \\
\twi{Q}_{\ss{A}} &=& {1\over1-v^2} \left[ (1-2v^2)2N + v^2 Q_{\ss{A}} \right]
\nn\\
\twi{Q}_{\ss{E}} &=& {1\over1-v^2} \left[ (1+v^2) Q_{\ss{E}} +2v (2 \GN
M - Q_{\ss{D}} ) \right] \nn\\
\twi{Q}_{\ss{M}} &=& Q_{\ss{M}} + {v\over1-v^2} \left[ 2N - Q_{\ss{A}} \right]
\ \ . \nn
\eeqa
It is assumed in these equations that $v^2<1$. Notice that the previous results
are obtained as $v \to 0$. The `tilde' is written over Newton's constant in the
first line to emphasize that this constant is modified because of the shift
which was required of the dilation field in order to preserve its boundary
conditions at large $r$.

We are now in a position to argue against the necessity of physical equivalence
between dual solutions when the duality which relates the solutions is based on
a noncompact symmetry. The first step is the observation that the asymptotic
value, $v$, which is taken by $A_t$ drops out of all physical quantities
because it can be changed by performing a gauge transformation. The
one-parameter family of field configurations which differ only in their
asymptotic value for $A_t$ are therefore all physically equivalent to one
another. In particular, all would agree on their prediction for the scattering
of an incident massless string mode.

Now, if duality based on eqs.~\pref{dualitytransformation} were always to
generate physically equivalent string configurations, it would follow that the
one-parameter family of dual configurations which are obtained from the original
gauge-equivalent family must also all be physically equivalent.  But
eqs.~\pref{longeq} show that the dual configurations have conserved charges
which depend explicitly on the value of the parameter, $v$. They therefore
differ in their physical predictions, such as for the scattering of massless
string modes, and so cannot be physically equivalent. We conclude that for the
class of configurations considered here, eqs.~\pref{dualitytransformation} are
{\em not} symmetries of string theory. 

Notice that the compactness of the symmetry plays a role in this argument. 
If the symmetry direction were compact, the above argument would fail since it
would no longer be possible to shift $A_t$ by an arbitrary constant value simply
by performing a gauge transformation. Different constant values for $A_t$ would
be physically distinct since they could be characterized by different values for
a gauge-invariant quantity: the Wilson lines around the compact direction.

\subsection{Flat Space Revisited}

The above considerations can be further focussed by reconsidering the case of
flat space and toroidal compactifications. We can avoid the complication of
understanding the meaning of winding and momentum modes in the time direction
by considering instead duality based on translations in one of the noncompact
spatial directions. It is instructive, in this case, to consider this direction
as the limit of a compact, circular, direction as its circumference tends to
infinity. 

For any finite circumference, duality indeed represents an exact string
symmetry, with all momentum modes dualizing to winding modes in the symmetry
direction, and {\em vice versa}. For infinite circumference the momentum states
degenerate into a continuum of permitted eigenvalues, while the masses of all of
the winding modes tend to infinity. As required by duality, precisely the
opposite happens as the circumference tends to zero: winding states form a
continuum and momentum states become infinitely massive. So long as both
winding- and momentum- modes are both kept in the infinite (and zero-)
circumference limit, duality continues to relate equivalent theories.

But string theory, as it is normally defined on (noncompact) flat Minkowski
space, is the path integral over all possible string embeddings in this space.
This corresponds to what is obtained from a compact space in the
infinite-circumference limit {\em provided} that the winding modes are dropped
in this limit. Once the winding modes are omitted, the equivalence of dual
configurations is no longer guaranteed.

This picture is similar to that obtained by Ref.~\cite{AABL}, who analyze more
systematically the equivalence of solutions related by duality based on a
noncompact symmetry. They find equivalence, but only if one of the dual pair of
solutions is quantized as a vortex gas. In this unconventional string theory,
there are no local vertex operators carrying momentum in the symmetry direction,
but instead one constructs nonlocal `vortex' operators carrying
`winding-number' for this direction. 

This line of argument indicates that a string theory is not uniquely specified
by listing its configuration of background fields, since additional choices ---
such as whether or not to include nonlocal vortex operators --- must also be
made. Although duality can always produce an equivalent string theory from any
given one, the two string theories generically will {\it not} both be
interpretable as the sum over string embeddings in the given background field
configurations. In comparing the dual theories, one must keep in mind this
distinction. For example, for two-dimensional black holes, the propagation and
interactions of tachyons near the singularity are equivalent to the propagation
and interactions near the horizon of some dual vortex states, rather than of
tachyons.

\section{Conclusions}

The purpose here has been to present some results of a preliminary investigation
\cite{bhsoln,bhdual} into the consequences of duality, and related,
transformations on our understanding of string propagation through
black-hole-like spacetimes.  This has been accomplished by outlining two
qualitatively different types of results.  

The first class of result simply uses these transformations to
generate new solutions to the string equations. This technique has been used
to generate the most general static, spherically-symmetric and
asymptotically-flat string configuration which solves the string equations to
lowest order in $\ap$. Solutions of this form are now known which involve all of
the metric, dilaton, axion and gauge fields --- a broader class than had
hitherto been constructed. They are labelled by a collection of five conserved
quantities: the mass $M$, and the dilaton, axion, electric and magnetic
charges: $Q_{\ss{D}}$, $Q_{\ss{A}}$, $Q_{\ss{E}}$ and $Q_{\ss{M}}$. If the
static condition is relaxed to include some stationary metrics, then a sixth
charge, the NUT parameter, $N$, is also required. 

Using duality, and related, transformations to generate these solutions has the
practical advantage of requiring the use of only algebraic techniques, instead
of having to solve several coupled, nonlinear, partial-differential equations.
This makes its application relatively straightforward to apply to more
complicated, less symmetric, field configurations. 

Most of the solutions which were generated in this way have real curvature
singularities at fixed, nonzero $r$, and these singularities are in some cases
naked. We regard it to be premature to discard these solutions until it is
better understood precisely what configurations string theory considers to be
singular. 

The second type of investigation reported here concerns the action and
implications of duality on these low-energy solutions in particular, and on
static, spherically-symmetric and asymptotically-flat string configurations in
general. 

The result can be presented in a very general form. By investigating the action
of duality on the asymptotic form of a general field configuration for large
$r$, it becomes clear that duality simply interchanges the configuration's mass
and dilaton charge, as well as interchanging the axion charge with the NUT
parameter. The electric and magnetic charges do not change. These results are
very robust --- applying equally well once higher orders in $\ap$ are included
--- depending as they do only on the large-$r$ behaviour of the solution. 

Somewhat surprisingly we find that the dual field configuration depends in an
important way on the assumed asymptotic form for the electrostatic potential,
$A_t$. The conclusions of the previous paragraph assumed $A_t$ to be chosen
to vanish asymptotically. 

For the special case of the four-dimensional `Schwarzschild-like'
solutions --- \ie\ those that are nonsingular at the Schwarzschild radius ---
a  black hole of mass $M$ is mapped by duality onto a singular configuration
whose mass, $\twi M$, is given by $\twi{M} = - k/(\ap M)$, where $k = \lambda
\ap^2 / 2 \GN^2 = 2\lambda e^{2\phi_0}$. $\lambda$ here is $\hf$ in the bosonic,
and $\nth{4}$ in the heterotic string. For the superstring, it happens that
$\lambda = k = 0$, and so $O(\ap^3)$ corrections are required in order to
determine the dual mass. The result in this case turns out to be: $\twi{M} = - 
\, k' / ( \ap^3 M^5)$, where  $k' = 3 \zeta(3) \, e^{6 \phi_0}$.

We have also addressed the potential inequivalence of the two dual solutions in
the case where the symmetry on which duality is based is not compact. An argument
was presented which led to one of the following two options: ($i$) dual
solutions, defined by eqs.~\pref{dualitytransformation}, can be physically
inequivalent; or ($ii$) constant shifts of the electrostatic potential, $A_t$,
can change the physical content of the theory. Arguments in favour of the first
option were presented, using the well-understood example of flat space. 

The potential implications of duality are just beginning to be explored, both
for black holes, and for other physical systems. May the rewards be both rich
and varied! 

\section{Acknowledgements}
 
C.B. would like to thank the organizers for their kind invitation to speak at
the conference. Our funds have been provided by N.S.E.R.C.\ of Canada, les Fonds
F.C.A.R.\ du Qu\'ebec and the Swiss National Foundation.

\def\pr#1#2#3{{\it Phys.~Rev.} {\bf #1} (19#2) #3}
\def\np#1#2#3{{\it Nucl.~Phys.} {\bf #1} (19#2) #3}
\def\pl#1#2#3{{\it Phys.~Lett.} {\bf #1} (19#2) #3}
\def\prc#1#2#3{{\it Phys.~Rev.} {\bf C#1} (19#2) #3}
\def\prd#1#2#3{{\it Phys.~Rev.} {\bf D#1} (19#2) #3}
\def\prl#1#2#3{{\it Phys. Rev. Lett.} {\bf #1} (19#2) #3}
\def\plb#1#2#3{{\it Phys. Lett.} {\bf B#1} (19#2) #3}
\def\npb#1#2#3{{\it Nucl. Phys.} {\bf B#1} (19#2) #3}
\def\mpla#1#2#3{{\it Mod.\ Phys.\ Lett.} {\bf A#1}, (19#2) #3}
\def\etal{{\it et.al. \/}}

\bibliographystyle{unsrt}

\end{document}